\documentclass[aps,prd,nofootinbib,preprint,tightenlines,showpacs]{revtex4}
\newcommand{\be}{\begin{equation}}
\newcommand{\ee}{\end{equation}}
\newcommand{\bea}{\begin{eqnarray}}
\newcommand{\eea}{\end{eqnarray}}
\def\bml{\begin{subequations}}
\def\eml{\end{subequations}}
\def\blea{\bml\bea}
\def\elea{\eea\eml}

\def\bx{\mathbf{x}}

\def\bx{\mathbf{x}}

\def\bz{\mathbf{z}}
\def\Re{\mathop{\rm Re}}
\def\Im{\mathop{\rm Im}}

\def\sgn{\mathop{\rm sgn}}
\def\Tsplit{T^{\text{split}}}
\def\Tren{T^{\text{ren}}}
\def\Vmax{V_{\text{max}}}
\usepackage{amsmath}
\usepackage{amsfonts}
\usepackage{latexsym}
\usepackage{epsfig}
\usepackage{epsfig}

\begin{document}

\title{Quantum inequality for a scalar field with a background
  potential}

\author{Eleni-Alexandra Kontou}
\author{Ken D. Olum}
\affiliation{Institute of Cosmology, Department of Physics and Astronomy,\\ 
Tufts University, Medford, MA 02155, USA}

\begin{abstract}

Quantum inequalities are bounds on negative time-averages of the
energy density of a quantum field.  They can be used to rule out
exotic spacetimes in general relativity.  We study quantum
inequalities for a scalar field with a background potential (i.e., a
mass that varies with spacetime position) in Minkowski space.  We
treat the potential as a perturbation and explicitly calculate the
first-order correction to a quantum inequality with an arbitrary
sampling function, using general results of Fewster and Smith.  For an
arbitrary potential, we give bounds on the correction in terms of the
maximum values of the potential and its first three derivatives.  The
techniques we develop here will also be applicable to quantum
inequalities in general spacetimes with small curvature, which are
necessary to rule out exotic phenomena.

\end{abstract}


\maketitle

\section{Introduction}

General Relativity relates spacetime curvature to the stress-energy
tensor $T_{ab}$, but does not provide any constraints on what $T_{ab}$
might be.  Thus relativity alone allows us to construct any spacetime,
including those with exotic features, such as wormholes and time
machines.  However in the context of quantum field theory, while
negative energies are possible, for example in the Casimir effect,
there are various constraints on the stress energy tensor. One example
is averaged energy conditions that provide bounds on integrals of
$T_{ab}$ along an entire geodesic.  Another example of bounding the
stress-energy tensor is quantum inequalities that bound the total
energy when averaging over a time period.

Quantum inequalities were introduced by Ford \cite{Ford:1978qya} to avoid
the possibility of violating the second law of thermodynamics by
sending a flux of negative energy into a black hole.  The general form
of a quantum inequality is
\be\label{QI}
\int_{-\infty}^\infty d\tau\,w(\tau) T_{ab}(x(\tau)) V^a V^b > - B\,,
\ee
where $x(\tau)$ is a timelike path parameterized by proper time $\tau$
with tangent vector $V$, and $w$ is a sampling function.  The quantity
$B$ is a bound, depending on the function $w$ and the quantum field of
interest.

Since the original work of Ford, quantum inequalities have been
derived for a wide range of different fields and sampling functions.
However, these quantum inequalities apply only to free fields in
Minkowski space without boundaries.  In other cases, there are
difference quantum inequalities \cite{Ford:1994bj}, in which $T_{ab}$
in Eq.~(\ref{QI}) is replaced by the difference between $T_{ab}$
in some state of interest and $T_{ab}$ in a reference state.  The
bound $B$ may also then depend on the reference state. However, such
difference inequalities cannot be used to rule out exotic spacetimes,
at least in the case where the exotic matter that supports the
spacetime comes from the vacuum state in the presence of the
boundaries.

Nevertheless Ref.~\cite{Fewster:2006uf} shows that boundaries do not
allow violation of the average null energy condition (ANEC), which
states that
\be
\int_{-\infty}^\infty d\lambda\,T_{ab}(\gamma(\lambda)) V^a V^b \ge 0\,,
\ee
where the integral is taken on a null geodesic $\gamma$, affinely
parameterized by $\lambda$ with tangent vector $V$.
Reference~\cite{Graham:2007va} proved that ANEC is sufficient to rule
out many exotic spacetimes.  The proof made use of quantum
inequalities for null contractions of the stress tensor averaged over
timelike geodesics \cite{Fewster:2002ne}.

None of this work, however, really addresses the possibility of exotic
spacetimes.  The quantum inequalities on which it depends apply only
in flat spacetime, so they cannot be used to rule out spacetimes with
exotic curvature.  For that, we need limits on the stress-energy
tensor in curved spacetimes.  One possible approach is to appeal to
the principle of equivalence to say that if the averaging timescale of
the quantum inequality is small compared to the curvature radius of
the spacetime, then flat-space results should apply approximately
\cite{Ford:1995wg}.  We used such reasoning in
Ref.~\cite{Kontou:2012ve} to conjecture that flat-space quantum
inequalities apply, even in curved space, with certain corrections,
which we hoped were not too large.  From this conjecture, we were able
to extend the argument of Ref.~\cite{Fewster:2006uf} to curved
spacetime.  But the truth of our conjecture is not known.

As a first step toward proving the conjecture of
Ref.~\cite{Kontou:2012ve}, we derive in the present work a quantum
inequality in a flat spacetime with a background potential, i.e.,
a field with a mass depending on spacetime position. This is a simpler
system that has many of the important features of quantum fields in
curved spacetime. For a scalar field $\Phi$ in a background
potential, the Lagrangian is
\be
L=\frac{1}{2} \left[\partial_\mu \Phi \partial^\mu \Phi-V(x) \Phi^2 \right]\,,
\ee
the equation of motion is
\be
(\Box+V(x)) \Phi=0\,,
\ee
and the classical energy density is
\be\label{classicalT00}
T_{00}=\frac12\left[(\partial_t \Phi)^2+(\nabla \Phi)^2+V(x)\Phi^2\right]\,.
\ee

We work only in first order in $V$ but don't otherwise assume that it is small. We can express the maximum values of the background potential and its derivatives as 
\bea\label{Vmax}
\begin{array}{cc}
|V| \leq \Vmax & |V_{,a}| \leq \Vmax' \\
|V_{,ab}| \leq \Vmax'' & |V_{,abc}| \leq \Vmax'''\,,
\end{array}
\eea
where $\Vmax$, $\Vmax'$, $\Vmax''$ and $\Vmax'''$ are positive
numbers, finite but not necessarily small.

Our proof uses a general absolute quantum inequality proven by Fewster
and Smith \cite{Fewster:2007rh}, which we discuss in
Sec.~\ref{sec:FSQEI}.  This inequality gives a bound on the
renormalized energy density based on the Fourier transform of the
point-split energy density operator applied to the Hadamard series.
In Sec.~\ref{sec:general}, we discuss this operator, in
Sec.~\ref{sec:H} we compute the Hadamard series, and in
Sec.~\ref{sec:TH}, we apply the operator.  In Sec.~\ref{sec:Fourier},
we perform the Fourier transform, leading to the final quantum
inequality in Sec.~\ref{sec:QI}.  We conclude in
Sec.~\ref{sec:conclusions} with a discussion of future possibilities.

We use metric signature $(+,-,-,-)$. Indices $a,b,c, \dots$ denote all
spacetime coordinates while $i,j,k \dots$ denote only spatial coordinates.

\section{Absolute Quantum Energy Inequality}
\label{sec:FSQEI}

We start by defining the renormalized energy density according to the
renormalization procedure of Wald \cite{Wald:qft}.  Let $\langle
\phi(x)\phi(x') \rangle$ be the two-point function of the scalar
field, and define the Hadamard form
\be \label{hadamard}
H(x,x')=\frac{1}{4\pi^2} \left[ \frac{1}{\sigma_+(x,x')}+\sum_{j=0}^{\infty}v_j(x,x') \sigma_+^j (x,x') \ln(\sigma_+(x,x'))+\sum_{j=0}^{\infty}w_j (x,x')\sigma^j (x,x') \right]\,,
\ee
where
\be
\sigma(x,x')=-\eta_{ab} (x-x')^a (x-x')^b\,,
\ee
so that $\sigma(x,x')<0$ when the separation between $x$ and $x'$ is
timelike. By $F(\sigma_+)$, for some function $F$, we mean the
distributional limit
\be 
F(\sigma_+)=\lim_{\epsilon \to 0^+} F(\sigma_{\epsilon})\,,
\ee
where
\be
\sigma_{\epsilon}(x,x')=\sigma(x,x')+2i \epsilon(t-t')+\epsilon^2\,,
\ee
with $t$ and $t'$ being the time components of the 4-vectors $x$ and
$x'$.
In most of the calculation we consider $x$ and $x'$ separated only
in time.  In that case, we define
\be
\tau=t-t'\,,
\ee
and write
$F(\tau_-)$ to mean $\lim_{\epsilon \to 0^+} F(\tau-i\epsilon)$.
In general the sums in Eq.~(\ref{hadamard}) do not converge, but we
will be concerned only with the first few terms.  Following Wald 
\cite{Wald:qft}, we will choose $w_0 = 0$.

When the scalar field is in a Hadamard state, the singularity
structure of $\langle\phi(x)\phi(x')\rangle$ is precisely that of
$H(x,x')$, so the renormalized two-point function
$\langle\phi(x)\phi(x')\rangle -H(x,x')$ is smooth.  To this we apply
a point-split energy density operator, which is analogous
to the classical energy density of Eq.~(\ref{classicalT00}),
\be\label{Tsplit}
\Tsplit=\frac{1}{2} \left[ \sum_{a=0}^3 \partial_a \partial_{a'}
 +\frac{V(x)+V(x')}2 \right]\,,
\ee
and take the limit where $x$ and $x'$ coincide.  In this limit, the
location of evaluation of $V$ does not matter, but the form above will
be convenient later.  Thus we define
\be \label{Tren}
\langle \Tren_{00}(x') \rangle \equiv \lim_{x\to x'}  \Tsplit \left( \langle \phi(x)\phi(x') \rangle-H(x,x') \right)-Q(x')\,,
\ee
where $Q$ is a term added ``by hand'' to prevent the failure of
conservation of the stress-energy tensor. Wald \cite{Wald:1978pj}
derived this term for curved spacetime. The calculation for flat space
with background potential is essentially the same, giving
\be \label{Q}
Q(x)=\frac{1}{12\pi^2}w_1(x,x)\,.
\ee

Unfortunately, there is an ambiguity in the above procedure.  In order
to take logarithms, we must divide $\sigma$ by the square of some
length scale $l$.  Changing the scale to some other scale $l'$
decreases $H$ by $\delta H =2 (v_0+v_1\sigma+\cdots)\ln(l'/l)$.  This
results in increasing $T_{ab}$ by $\lim_{x\to x'}
(\partial_a\partial_b- (1/2) \eta_{ab}\partial^c\partial_c)\delta H$.
Using the values for $v_0$ and $v_1$ computed below, this becomes
$(1/12) (V_{ab} -\eta_{ab}\Box V)\ln(l'/l)$.  Thus we see that
the definition of $T_{ab}$ must include arbitrary multiple of $(V_{ab}
-\eta_{ab}\Box V)$.  This ambiguity can also be understood as the
possibility of including in the Lagrangian density a term of the form
$R(x) V(x)$, where $R$ is the scalar curvature.  Varying the metric to
obtain $T_{ab}$ and then going to flat space yields the above term.
The situation is very much analogous to the possible addition of terms
of the form $R^2$ and $R_{ab} R^{ab}$ in the case of a field in curved
spacetime.

Thus we rewrite Eq.~(\ref{Tren}) to include the ambiguous term,
\be
\langle \Tren_{00}(x') \rangle \equiv \lim_{x\to x'}  \Tsplit \left(
\langle \phi(x)\phi(x') \rangle-H(x,x') \right)-Q(x')+CV_{,ii}\,,
\ee
where $C$ is some constant.  Whenever definition of $T_{00}$ one is
trying to use, one can pick an arbitrary scale $l$ and adjust $C$
accordingly.

Now, following Ref.~\cite{Fewster:2007rh} we define
\be\label{Htilde}
\tilde H(x,x')=\frac{1}{2} \left[ H(x,x')+H(x',x)+iE(x,x') \right]\,,
\ee
where $E$ is the advanced-minus-retarded Green's function, and thus
$iE$ is the antisymmetric part of the two point function.  We use the Fourier
transform convention
\be\label{Fourier}
\hat{f}(k) \text{ or } [f]^{\wedge}(k)=\int_{-\infty}^\infty dx\,f(x) e^{ixk}\,.
\ee
We consider the energy density integrated along a geodesic on the $t$
axis with a smooth, positive sampling function $g(t)$.  The absolute
quantum inequality of Ref.~\cite{Fewster:2007rh} for this case is
\be\label{qinequality}
\int_{-\infty}^\infty d\tau\,g(t)^2 \langle \Tren_{00} \rangle (t,0) 
\geq -B\,,
\ee 
where
\be\label{B}
B = \int_0^\infty\frac{d\xi}{\pi}\hat F(-\xi,\xi)
+\int_{-\infty}^\infty dt\,g^2(t) (Q-CV_{,ii})\,,
\ee
and
\be\label{F}
F(t,t') = g(t)g(t')\Tsplit \tilde H_{(5)}((t,0),(t',0))\,,
\ee
$\hat F$ denotes the Fourier transform in both arguments according to
Eq.~(\ref{Fourier}), and the subscript $(5)$ means that we include
only terms through $j = 5$ in the sums of Eq.~(\ref{hadamard}).

\section{General considerations}
\label{sec:general}
\subsection{Smooth, symmetrical contributions}
\label{sec:symmetry}

Let $\bar{x}=\frac{x-x'}{2}$, $\bar t = (t+t')/2$ and $\tau = t-t'$.  Let
\be\label{AF}
A(\tau) = \int_{-\infty}^\infty d\bar t\,F\left(\bar t+\frac{\tau}{2},\bar t-\frac{\tau}{2}\right)\,.
\ee
Then $\hat F(-\xi,\xi) = \hat A(-\xi)$.

Suppose $F$ contains some term $f$ that is symmetrical in $t$ and
$t'$.  Let $a$ be the corresponding term in $A$ according to
Eq.~(\ref{AF}).  Then $a$ will be even in $\tau$, so $\hat a$ will
be even also.  If $a\in C^1$, then $\hat a\in L^2$, and we can perform
the integral of this term separately, giving an inverse Fourier
transform,
\be
\int_0^\infty\frac{d\xi}{\pi}\hat f(-\xi,\xi)
= \int_{-\infty}^\infty\frac{d\xi}{2\pi} \hat a(\xi) = a(0)\,.
\ee
In particular, if
\be
\lim_{t\to t'} f(t,t') = f(t)\,,
\ee
then
\be
\int_0^\infty\frac{d\xi}{\pi}\hat f(-\xi,\xi)
= \int_{-\infty}^\infty dt\,g(t)^2 f(t)\,,
\ee 
and if $f(t)=0$ there is no contribution.  

Terms arising from $H$ appear symmetrically in $\tilde H$.  At orders
$j>1$ they have at least 4 powers of $\tau$, so they vanish in the
coincidence limit even when differentiated twice by the operators of
$\Tsplit$.  Thus such terms make no contribution to
Eq.~(\ref{B}).

\subsection{Simplification of $\Tsplit$}
We would like to write the operator $\Tsplit$ in terms of separate
derivatives on the centerpoint $\bar x$ and the difference between the
points.  First we separate the derivatives in $\Tsplit$ into time and
space,
\be \label{spti}
\sum_{a=0}^3 \partial_a \partial_{a'}=\partial_t
\partial_{t'}+\nabla_x \cdot \nabla_{x'}\,.
\ee
We can expand the spatial derivative with respect to\footnote{When a
  derivative is with respect to $x$ or $x'$, we mean to keep the other
  of these fixed, while when the derivative is with respect to $\bar t$
  or $\tau$, we mean to keep the other of these fixed.}  $\bar x$,
\be\label{barxd}
\nabla_{\bar{x}}^2=\nabla_x^2+2\nabla_x\cdot \nabla_{x'}+\nabla_{x'}^2\,,
\ee
and Eqs.~(\ref{Tsplit},\ref{spti},\ref{barxd}) give
\bea
\Tsplit&=&\frac{1}{2}\left[\partial_t \partial_{t'}+\frac{1}{2}
  \left(\nabla_{\bar{x}}^2-\nabla_x^2-\nabla_{x'}^2\right)
+\frac12\left(V(x)+V(x')\right)\right]=\nonumber\\
&=&\frac{1}{4}\left[\nabla_{\bar{x}}^2+\Box_x-\partial_t^2
+\Box_{x'}-\partial_{t'}^2+2\partial_t\partial_{t'}+V(x)+V(x')\right]\,,
\eea
where $\Box_x$ and $\Box_{x'}$ denote the D'Alembertian operator with
respect to $x$ and $x'$.  Then using
\be\label{bartd}
\partial_\tau^2=\frac14\left[\partial_t^2-2\partial_t \partial_{t'}+\partial_{t'}^2\right]\,,
\ee
we can write
\be
\Tsplit \tilde{H}=\frac{1}{4}\left[\left( \Box_x+V(x) \right) \tilde{H}
+\left( \Box_{x'}+V(x') \right)\tilde{H}+\nabla_{\bar{x}}^2 \tilde{H}\right]-\partial_\tau^2 \tilde{H}\,.
\ee
Consider the first term.
The function $H(x,x')$ obeys the equation of motion in $x$, and so
does $E(x,x')$.  Thus
\be \label{eqmd}
\left( \Box_x+V(x) \right) \tilde{H}=\frac{1}{2}(\Box_x+V) H(x',x)\,.
\ee
The only asymmetrical part of $H$ comes from the $w_j$, so 
\be
H(x',x) = H(x,x') + \frac{1}{4\pi^2}\sum_j(w_j(x',x)-w_j(x,x'))\sigma^j(x,x')\,.
\ee
Terms involving both $V$ and $w_j$ are second order in $V$, so we
can ignore them, giving
\be\label{EofMH1}
\left( \Box_x+V(x) \right) \tilde{H}
= \frac{1}{4\pi^2}\Box_x\sum_j(w_j(x',x)-w_j(x,x'))\sigma^j(x,x')\,.
\ee
Similarly,
\be\label{EofMH2}
\left( \Box_{x'}+V(x) \right) \tilde{H}
= \frac{1}{4\pi^2}\Box_{x'}\sum_j(w_j(x,x')-w_j(x',x))\sigma^j(x,x')\,.
\ee
Adding together Eqs.~(\ref{EofMH1},\ref{EofMH2}), we get something
which is symmetric in $x$ and $x'$ and vanishes in the coincidence
limit.  Thus according to the analysis of Sec.~\ref{sec:symmetry}, it
makes no contribution and for our purposes we can take
\be \label{T}
\Tsplit \tilde{H}=\left[\frac{1}{4}\nabla_{\bar{x}}^2-\partial_\tau^2
\right] \tilde{H}\,.
\ee

\section{Computation of $\tilde H$}
\label{sec:H}

Examining Eq.~(\ref{T}) we see that is sufficient to compute $\tilde
H$ for purely temporal separation as a function of $t$, $t'$, and
$\bx$, the common spatial position of the points.  The function
$H(t,t')$ is a series of terms with decreasing degree of singularity
at coincidence: $\tau^{-2}$, $\ln\tau$, $\tau^2\ln\tau$, etc.  For the
first term in Eq.~(\ref{T}), terms in $H$ that have any positive
powers of $\tau$ will not contribute by the analysis of
Sec.~\ref{sec:symmetry}.  For the second term we need to keep terms in
$H$ up to order $\tau^2$, because the derivatives will reduce the
order by 2.

The symmetrical combination $H(t,t') + H(t,t')$, will lead to
something whose Fourier transform does not decline rapidly for
positive $\xi$, so that if this alone were put into
Eq.~(\ref{B}) the integral over $\xi$ would not converge.
But each term in $H(t,t') + H(t,t')$ will combine with a term coming
from $iE(x,x')$ to give something whose Fourier transform does decline
rapidly.

We will work order by order in $\tau = t - t'$ and write $H_j(t,t')$,
$j=-1,0,1,\ldots$, to denote the term in $H$ involving $\tau^{2j}$
(with or without $\ln\tau$), and $H_{(j)}$ to to denote the sum of all
terms up through $H_j$ .  We will split up $E(x,x')$ into terms
labeled $E_j$ that are proportional to $\tau^{2j}$, define a
``remainder term''
\be
R_j = E - \sum_{k=-1}^j E_k\,,
\ee
and let
\blea
\tilde
H_{j}(x,x') &=& \frac12\left[H_j(x,x') + H_j(x',x) + iE_j(x,x'))\right]\\
\tilde H_{(j)}(x,x') &=& \frac12\left[H_{(j)}(x,x') + H_{(j)}(x',x) + iE(x,x'))\right]\,.
\elea

\subsection{General computation of $E$}

We will need the Green's functions for the background potential,
including only first order in $V$, so we write
\be
G=G^{(0)}+G^{(1)}+ \cdots.
\ee
The equation of motion is
\be \label{eqm}
(\Box+V(x)) G(x,x')=\delta^{(4)}(x-x')\,.
\ee
Using $\Box G^{(0)}(x,x')=\delta^{(4)}(x,x')$ and keeping only first-order
terms we have
\be
\Box G^{(1)}(x,x')=-V(x)G^{(0)}(x,x')\,,
\ee
so
\be
G^{(1)}(x,x')=-\int d^4 x'' G^{(0)}(x,x'') V(x'') G^{(0)}(x'',x')\,.
\ee
For $t>t''>t'$ we have for the retarded Green's function,
\be
G_R^{(0)}(x'',x')=\frac{1}{2\pi} \delta((t''-t')^2-|\bx''-\bx'|^2)=\frac{1}{4\pi} \frac{\delta(t''-t'-|\bx''-\bx'|)}{|\bx''-\bx'|}\,.
\ee
So we can write
\be
G_R^{(1)}(x,x')=- \frac{1}{8\pi^2} \int d^3 \bx'' \int dt''\delta((t-t'')^2-|\bx-\bx''|^2) \frac{\delta(t''-t'-|\bx''-\bx'|)}{|\bx''-\bx'|} V(t'',\bx'')\,.
\ee
Integrating over the second delta function we find
$t''=t'+|\bx''-\bx'|$.  Again considering purely temporal separation and
defining $\bz''=\bx''-\bx'$ and $z'' = |\bz''|$, we find
\be
G_R^{(1)}(t,t')=-\frac{1}{8\pi^2}\int d\Omega \int dz'' z''^2  \frac{\delta(\tau^2-2\tau z'')}{z''} V(t'+z'',\bx'+z'' \hat{\Omega})\,,
\ee 
where $\int d\Omega$ denotes integration over solid angle, and
$\hat{\Omega}$ varies over all unit vectors.  We can integrate over
$z''$ to get $z''=\tau/2$ and
\be
G_R^{(1)}(t,t')= - \frac{1}{32\pi^2}\int d\Omega\,V(\bar{t},\bx'+\frac{\tau}{2}\hat{\Omega})\,.
\ee
If we define a 4-vector $\Omega=(0,\hat{\Omega})$ we can write
\be \label{G1}
G_R^{(1)}(t,t')=-\frac{1}{32 \pi^2} \int d\Omega\,V(\bar{x}+\frac{\tau}{2} \Omega)\,.
\ee
The advanced Green's functions are the same with $t$ and $t'$ reversed.
Since $E$ is the advanced minus the retarded function, we have
\be\label{Egeneral}
E^{(1)}(t,t') = \frac{1}{32 \pi^2} \int d\Omega\,V(\bar{x}+\frac{|\tau|}{2} \Omega)\sgn\tau\,.
\ee

\subsection{Terms of order $\tau^{-2}$}

We now compute the various $H_j$, $\tilde H_j$, and $E_j$, starting
with terms that go as $\sigma^{-1}$ or $\tau^{-2}$.  These terms are
exactly what one would have for flat space without potential.
Equation~(\ref{hadamard}) gives
\be
H_{-1}(x,x')=\frac{1}{4\pi^2\sigma_+(x,x')}=-\frac{1}{4\pi^2(\tau_-^2-z^2)}
\,,
\ee
where
\be
\bz=\bx-\bx'
\ee
and
\be
z = |\bz|\,.
\ee

Similarly, the advanced minus retarded Green's function to this order is
\be
E_{-1}(x,x') = G_A(x,x') - G_R(x,x')
= \frac{\delta(\tau-z) - \delta(\tau+z)}{4\pi z}\,,
\ee
so
\be
\tilde H_{-1}(t,t') = \lim_{z \to 0} \frac{1}{8\pi^2}\left[
-\frac{1}{\tau_+^2-z^2}-\frac{1}{\tau_-^2-z^2}+i \pi
\frac{\delta(\tau+z)-\delta(\tau-z)}{z} \right]\,,
\ee
where 
\be
F(\tau_+)=\lim_{\epsilon \to 0} F(\tau+i\epsilon)\,.
\ee
Taking the $\epsilon\to0$ limit in $\tau_+$ and $\tau_-$ gives the
formula
\be
-\frac{1}{\tau_+^2-z^2}+\frac{1}{\tau_-^2-z^2}=-i \pi \frac{\delta(\tau+z)-\delta(\tau-z)}{z} 
\ee
so
\be\label{H-1}
\tilde H_{-1}(t,t') =-\frac{1}{4\pi^2\tau_-^2} = H_{-1}(t,t')
\ee
as discussed in Ref~\cite{Fewster:2007rh}.

\subsection{Terms with no powers of $\tau$}

The Hadamard coefficients are given by the Hadamard recursion
relations, which are the solutions to
$(\Box+V(x''))H(x'',x')=0$, giving
\bml\label{vde}\be\label{v0de}
V(x'')+2 \eta^{ab} v_{0,a} \sigma_{,b}+4v_0+v_0\Box\sigma=0
\ee
\be\label{vj}
(\Box+V(x''))v_j+2(j+1)\eta^{ab}v_{j+1,a} \sigma_{,b}-4j(j+1)v_{j+1}+(j+1)v_{j+1}\Box\sigma=0\,.
\ee\eml
In Eqs.~(\ref{vde}), $\sigma_j$, $v_j$ and their derivatives are
functions of $x''$ and $x'$, and all derivatives act on $x''$.

To find the zeroth order of the Hadamard series we need only $v_0$.
For flat space, $\sigma_{,a}=-2\eta_{ab}(x''-x')^b$ and
$\Box{\sigma}=-8$.  Putting these in Eq.~(\ref{v0de})
 we have
\be \label{v01}
(x''-x')^a v_{0,a} + v_0 = \frac{V(x'')}{4}\,,
\ee
Now let $x'' = x' + \lambda(x-x')$ to integrate along the geodesic going
from $x'$ to $x$.  We observe that
\be
\frac{dv_0(x'',x')}{d\lambda} = (x-x')^a v_{0,a}(x'',x')\,.
\ee
So Eq.~(\ref{v01}) gives
\be
\lambda\frac{ dv_0(x'',x')}{d\lambda} + v_0(x'',x') = \frac{V(x'')}{4}\,,
\ee
or 
\be
\frac{d(\lambda v_0(x'',x'))}{ d \lambda} = \frac{V(x'')}{4}\,,
\ee
from which we immediately find
\be \label{v0}
v_0(x,x') = \int_0^1 d\lambda \frac{V(x' + \lambda(x-x'))}{4}\,.
\ee
Now we consider purely temporal separation so the background potential
is evaluated at $(t' + \lambda \tau, \bx)$.  We expand $V$ in a
Taylor series in $\tau$ around $0$ with $\bar t$ fixed,
\be\label{VTaylor}
V(t' + \lambda \tau)=V(\bar{t})+\tau (\lambda-\frac{1}{2})V_{,t} (\bar{t})+\frac{\tau^2}{2}(\lambda-\frac{1}{2})^2V_{,tt}(\bar{t})+\cdots.
\ee
We are calculating the zeroth order so we keep only the first term of
Eq.~(\ref{VTaylor}), and Eq.~(\ref{v0}) gives
\be
v_0(t,t')=\frac{1}{4}V(\bar{t})+O(\tau^2)
\ee
and thus
\be
H_0(x,x')=\frac{1}{16\pi^2} V(\bar{x}) \ln{(-\tau_-^2)}\,,
\ee
and
\be\label{H02}
H_0(x,x')+H_0(x',x)=  \frac{1}{4\pi^2}V(\bar{x}) \ln{|\tau|}\,.
\ee

We can expand $V$ around $\bar{x}$,
\be 
V(\bar{x}+\frac{\tau}{2}\Omega)=V(\bar{x})+V^{(1)} (\bar{x}+\frac{\tau}{2} \Omega)\,,
\ee
where $V^{(1)}$ is the remainder of the Taylor series
\be\label{V1}
V^{(1)}(\bar{x}+\frac{\tau}{2} \Omega)=V(\bar{x}+\frac{\tau}{2} \Omega) - V(\bar{x})
= \int_0^{\tau/2} dr\,V_{,i}(\bar{x}+r\Omega) \Omega^i\,.
\ee
Then from Eq.~(\ref{Egeneral}),
\blea
E_0(x,x')&=&\frac{1}{8\pi} V(\bar{x})\sgn\tau\label{E0}\\
R_0(x,x')&=&\frac{1}{32\pi^2}\int d\Omega\,V^{(1)} (\bar{x}+\frac{|\tau|}{2} \Omega)\sgn{\tau}\label{R0}\,.
\elea
Using
\be\label{lnsgn}
2\ln{|\tau|}+\pi i \sgn{\tau}= \ln{(-\tau_-^2)}\,,
\ee
we combine Eqs.~(\ref{H02},\ref{E0}) to find
\be\label{Ht0}
\tilde H_0(t,t')=\frac{1}{16\pi^2} V(\bar{x})\ln{(-\tau_-^2)}\,.
\ee
Combining all terms through order $0$ gives
\be\label{Ht-10}
\tilde H_{(0)}(t,t')=\tilde H_{-1}(t,t')+\tilde H_0(t,t')+\frac12i
R_0(t,t')\,.
\ee

\subsection{Terms of order $\tau^2$}

Now we compute the terms of order $\tau^2$ in $H$ and $E$.
First we need $v_0$ at this order, so we use Eq.~(\ref{VTaylor}) in
Eq.~(\ref{v0}), to get
\be\label{v02}
v_0(x,x')= \frac{1}{4}V(\bar{x})+\tau^2 \frac{1}{96} V_{,tt} (\bar{x})+\cdots\,.
\ee

Next we need to know $v_1$, but since $v_1$ is multiplied by $\tau^2$
in $H$, we need only the $\tau$-independent term $v_1(x,x)$.  From
Eq.~(\ref{vj}),
\be  \label{recv1}
(\Box+V(x))v_0(x,x')+2\eta^{ab}v_{1,a}(x,x')
\sigma_{,b}(x,x')+v_1(x,x')\Box_{x}\sigma(x,x') =0\,.
\ee
We neglect the $V(x)v_0$ term because it is second order in $V$.  At
$x=x'$, $\sigma_{,b} = 0$, so
\be\label{v1v0}
v_1(x,x)=\frac{1}{8}\lim_{x'\to x}\Box_x v_0(x,x')\,.
\ee
Using Eq.~(\ref{v0}) we find
\be
\Box_x v_0(x,x')=\frac{1}{4}\int_0^1 d\lambda \Box_x V(x' + \lambda(x-x'))=\frac{1}{4}\int_0^1 d\lambda\,\lambda^2 (\Box V)(x' + \lambda(x-x'))\,,
\ee
and Eq.~(\ref{v1v0}) gives
\be\label{v1}
v_1(x,x) =  \frac{1}{96}\Box V(\bar{x})\,.
\ee

We also need to know $w_1$, but again only at coincidence.
Reference~\cite{Wald:1978pj} gives
\be\label{w1}
w_1(x,x)=-\frac{3}{2} v_1(x,x)= -\frac{1}{64}\Box V(x)\,.
\ee
Combining the second term of Eq.~(\ref{v02}) with
Eqs.~(\ref{v1},\ref{w1}) gives
\be
H_1(t,t')=\frac{\tau^2 }{128\pi^2}
\left[\frac{1}{3} V_{,ii}(\bar{x})\ln{(-\tau_-^2)}
+\frac{1}{2}\Box V(\bar{x}) \right]\,.
\ee
Then $H_1(x',x)$ is given by symmetry, so
\be\label{H12}
H_1(x,x')+H_1(x',x)= \frac{\tau^2 }{64\pi^2} \left[\frac{2}{3} V_{,ii}(\bar{x}) \ln{|\tau|}+ \frac{1}{2}\Box V(\bar{x})\right]\,.
\ee

The calculation of $E_1$ is similar to that of $E_0$, but
now we have to include more terms in the Taylor expansion of $V$
around $\bar{x}$. So we expand
\be 
V(\bar{x}+\frac{\tau}{2}\Omega)=V(\bar{x})+\frac{1}{2}V_{,i}(\bar{x})\Omega^i\tau+\frac{1}{8}V_{,ij}(\bar{x})\Omega^i \Omega^j \tau^2+V^{(3)} (\bar{x}+\frac{\tau}{2} \Omega)\,,
\ee
where the remainder of the Taylor series $V^{(3)}$ is
\be\label{V3}
V^{(3)}(\bar{x}+\frac{\tau}{2} \Omega)=\frac12\int_0^{\tau/2} dr\,V_{,ijk}(\bar{x}+r\Omega)\left(\frac{\tau}{2}-r \right)^2\Omega^i \Omega^j \Omega^k dr\,.
\ee
Since $\int d\Omega\,\Omega^i = 0$ and
$\int d\Omega\,\Omega^i \Omega^j =(4\pi/3) \delta^{ij}$,
Eq.~(\ref{Egeneral}) gives
\blea
E_1(x,x')&=& \frac{1}{192 \pi} V_{,ii}(\bar{x})\tau^2\sgn\tau\label{E1}\,,\\
R_1(x,x')&=& \frac{1}{32\pi^2} \int d\Omega\,V^{(3)}
(\bar{x}+\frac{|\tau|}{2} \Omega)\sgn{\tau}\label{R1}\,.
\elea
Again using Eq.~(\ref{lnsgn}), we combine Eqs.~(\ref{H12},\ref{E1}) to get
\be\label{Ht1}
\tilde H_1(x,x')=\frac{\tau^2}{128\pi^2} \left[\frac{1}{3} \ln{(-\tau_-^2)} V_{,ii}+\frac{1}{2}\Box V(\bar{x}) \right]\,.
\ee
Combining all terms through order 1 gives
\be\label{Ht-11}
\tilde H_{(1)}(t,t')=\tilde H_{-1}(t,t')+\tilde H_0(t,t')+\tilde H_1(t,t')+\frac12
iR_1(t,t')\,.
\ee

\section{The $\Tsplit\tilde H$}
\label{sec:TH}

Using Eqs.~(\ref{F},\ref{T}), we need to compute
\be
\int_0^\infty \frac{d\xi}{\pi}\hat F(-\xi,\xi')\,,
\ee
where
\be
F(t,t') = g(t) g(t')\left[\frac{1}{4}\nabla_{\bar{x}}^2\tilde H_{(0)}(t,t')
-\partial_\tau^2  \tilde H_{(1)}(t,t')\right]\,.
\ee
Using
Eqs.~(\ref{H-1},\ref{R0},\ref{Ht0},\ref{Ht-10},\ref{R1},\ref{Ht1},\ref{Ht-11})
we can write this
\be
F(t,t') = g(t) g(t')\sum_{i=1}^6 f_i(t,t')\,,
\ee
with
\blea
f_1&=&\frac{3}{2\pi^2\tau_-^4}\\\
f_2&=&\frac{1}{8\pi^2\tau_-^2}V(\bar{x})\\
f_3&=&\frac{1}{96\pi^2} V_{,ii}(\bar{x}) \ln{(-\tau_-^2)}\\
f_4&=&-\frac{1}{128\pi^2}\left[V_{,tt}(\bar{x})+V_{,ii}(\bar{x})\right]\\
f_5&=&\frac{1}{256\pi^2} \int d\Omega\,\nabla_{\bar{x}}^2 \left[
  V^{(1)} (\bar{x}+\frac{|\tau|}{2} \Omega)\right]  i\sgn{\tau}\label{f5}\\
f_6&=& -\frac{1}{64\pi^2} \int d\Omega\,\partial_{\tau}^2\left[ V^{(3)} (\bar{x}+\frac{|\tau|}{2} \Omega)  i\sgn{\tau} \right]\,.\label{f6}
\elea

\section{The Fourier transform}
\label{sec:Fourier}

We want to calculate the quantum inequality bound $B$, given
by Eq.~(\ref{B}).  We can write it
\be
B=\sum_{i=1}^8 B_i\,,
\ee
where
\blea
B_i&=&\int_0^\infty  \frac{d\xi}{\pi} \int_{-\infty}^\infty dt
\int_{-\infty}^\infty dt' g(t) g(t') f_i(t,t') e^{i\xi(t'-t)}\nonumber\\
&=&\int_0^\infty \frac{d\xi}{\pi} \int_{-\infty}^\infty d\tau
\int_{-\infty}^\infty d\bar{t}\,
g(\bar{t}-\frac{\tau}{2})g(\bar{t}+\frac{\tau}{2})f_i(\bar t,\tau)e^{-i\xi \tau}
\qquad i = 1\ldots6\\
\label{B7}
B_7 &=& \int_{-\infty}^\infty dt\,g^2(t) Q(t)
= -\frac{1}{768\pi^2}\int_{-\infty}^\infty dt\,g^2(t)\Box V(t)\\
\label{B8}B_8 &=& -\int_{-\infty}^\infty dt\,g^2(t)CV_{,ii} (t)\,,
\elea
using Eqs.~(\ref{Q},\ref{B},\ref{w1}).

\subsection{The singular terms}

For $i=1,2,3$, $f_i$ consists of a singular function of $\tau$ times a
function of $\bar t$ (or a constant).  So we will separate the
singular part by writing
\be
f_i(\bar t,\tau) = g_i(\bar t) s_i(\tau)\,.
\ee
Then we define
\be
G_i(\tau)=\int_{-\infty}^\infty d\bar{t}\, g_i(\bar{t}) g(\bar{t}-\frac{\tau}{2})g(\bar{t}+\frac{\tau}{2})\,,
\ee
so
\be\label{Bi1}
B_i=\int_0^\infty \frac{d\xi}{\pi} \int_{-\infty}^\infty d\tau\,G_i(\tau) s_i(\tau) e^{-i\xi\tau}\,.
\ee
This is a Fourier transform of a product, so we can write it as a
convolution.  The $G_i$ are all real, even functions, and thus
their Fourier transforms are also, and we have
\be
B_i=\frac{1}{2\pi^2}\int_0^{\infty} d \xi\int_{-\infty}^{\infty} d\zeta\,\hat{G_i}(\xi+\zeta)\hat{s_i}(\zeta)\,.
\ee
Now if we change the order of integrals we can perform another change of variables $\eta=\xi+\zeta$, so we have
\be \label{general}
B_i=\frac{1}{2\pi^2}\int_{-\infty}^{\infty} d\zeta\int_\zeta^{-\infty} d\eta\,\hat{G_i}(\eta)\hat{s_i}(\zeta)
=\frac{1}{2\pi^2}\int_{-\infty}^{\infty}  d\eta\hat{G_i}(\eta) h_i(\eta)\,,
\ee
where
\be\label{hi}
h_i(\eta)=\int_{-\infty}^{\eta}d\zeta\,\hat{s_i}(\zeta)\,.
\ee
The arguments of Ref.~\cite{Fewster:2007rh} show that the integrals
over $\xi$ in Eq.~(\ref{Bi1}) and $\eta$ in Eq.~(\ref{hi}) converge.

We now calculate the Fourier transforms in turn, starting with $B_1$.  We have
\blea
g_1(\bar{t})&=&\frac{3}{2\pi^2} \\
s_1(\tau)&=&\frac{1}{\tau_-^4}\,.
\elea
The Fourier transform of $s_1$ is \cite{Gelfand:functions}
\be
\hat{s_1}(\zeta)=\frac{\pi}{3} \zeta^3 \Theta(\zeta)\,,
\ee
so
\be
h_1(\eta)=\int_0^{\eta} d\zeta\,\frac{\pi}{3}\zeta^3\Theta(\eta) =\frac{\pi}{12} \eta^4\Theta(\eta)\,.
\ee
From Eq.~(\ref{general}) we have
\be
B_1=\frac{1}{24\pi} \int_{0}^{\infty} d\eta\,\hat{G_1}(\eta) \eta^4\,.
\ee
Using $\widehat{f'}(\xi)=i \xi \hat{f}(\xi)$, we get
\be
B_1=\frac{1}{24\pi} \int_0^{\infty} d\eta\,\widehat{G_1''''}(\eta)\,.
\ee
The function $G_1$ is even, so its Fourier transform is also
even and we can extend the integral
\be
B_1=\frac{1}{48\pi} \int_{-\infty}^{\infty} d\eta\,\widehat{G_1''''}(\eta)=\frac{1}{24} G_1''''(0)\,.
\ee
For $G_1$ we have
\be
G_1(\tau)=\frac{3}{2\pi^2} \int d\bar{t}\, g(\bar{t}-\frac{\tau}{2})
g(\bar{t}+\frac{\tau}{2})\,,
\ee
and taking the derivatives and integrating by parts gives
\be \label{B1}
B_1=\frac{1}{16\pi^2} \int_{-\infty}^{\infty} d\bar{t}\, g''(\bar{t})^2\,,
\ee
reproducing a result of Ref.~\cite{Fewster:2007rh}.

For $B_2$ we have
\blea
g_2(\bar{t})&=&\frac{1}{8\pi^2}V(\bar{t}) \\
s_2(\tau)&=&\frac{1}{\tau_-^2}\,.
\elea
This calculation is the same as before except the Fourier transform of $s_2$ is
\cite{Gelfand:functions}
\be
\hat{s_2}(\zeta)=2\pi \zeta \Theta(\zeta)\,.
\ee
So we have
\be
B_2=-\frac{1}{2} G_2''(0)\,,
\ee
where
\be
G_2(\tau)=\frac{1}{8\pi^2} \int_{-\infty}^\infty d\bar{t}\, V(\bar{t}) g(\bar{t}-\frac{\tau}{2}) g(\bar{t}+\frac{\tau}{2})\,.
\ee
After taking the derivatives
\be \label{B2}
B_2=-\frac{1}{32\pi^2} \int_{-\infty}^\infty d\bar{t}\, V(\bar{t})
[g(\bar{t}) g''(\bar{t}) - g'(\bar{t}) g'(\bar{t})]\,.
\ee

For $B_3$ we have
\bea
s_3(\tau)&=&\ln(-\tau_-^2)\,.
\eea
In the appendix, we find the Fourier transform of $s_3$ as a distribution,
\be\label{s3hat}
\hat s_3[f] = 4\pi\int_0^\infty dk\,f'(k) \ln|k| -4\pi \gamma f(0)\,.
\ee
From Eq.~(\ref{hi}), we can write
\be
h_3(\eta)=\int_{-\infty}^\infty d\zeta\,\hat{s_3}(\zeta)\Theta(\eta-\zeta)\,,
\ee
which is given by Eq.~(\ref{s3hat}) with $f(\zeta)
=\Theta(\eta-\zeta)$, so
\be
h_3(\eta)=-4\pi\int_0^\infty d\zeta\,\delta(\eta-\zeta) \ln|\zeta| - 4\pi\gamma \Theta(\eta)=
-4\pi\Theta(\eta) ( \ln\eta + \gamma )\,.
\ee
Then Eq.~(\ref{general}) gives
\be
B_3= -\frac{2}{\pi}\int_0^\infty d\eta\,\hat{G_3}(\eta)
 \left( \ln\eta + \gamma \right)
= -\frac{1}{\pi}\int_{-\infty}^\infty d\eta\,\hat{G_3}(\eta)
( \ln|\eta| + \gamma )\,,
\ee
since $G_3$ is even.  The integral is just the distribution $w$ of
Eq.~(\ref{distw}) applied to $\hat G_3$, which is by definition $\hat
w[G_3]$, so Eq.~(\ref{distwhat}) gives
\be
B_3=-\int_{-\infty}^{\infty} d\tau\,G_3'(\tau)\ln{|\tau|}\sgn{\tau}\,,
\ee
with 
\be
G_3(\tau)=\frac{1}{96\pi^2}\int_{-\infty}^\infty dt
V_{,ii}(\bar{t}) g(\bar{t}-\frac{\tau}{2})g(\bar{t}+\frac{\tau}{2})\,,
\ee
so
\be\label{B3}
B_3 = -\frac{1}{48\pi^2}\int_{-\infty}^{\infty} d\tau\,\ln{|\tau|}\sgn{\tau}
\int_{-\infty}^{\infty} d\bar t\,V_{,ii}(\bar{t})
g(\bar{t}-\frac{\tau}{2})g'(\bar{t}+\frac{\tau}{2})\,.
\ee

\subsection{The non-singular terms}

For $i=4,5,6$, $f_i$ is not singular at $\tau=0$.  We include everything in
\be\label{Fi}
F_i(\tau)=\int_{-\infty}^\infty d\bar{t}\, f_i(\tau,\bar{t}) g(\bar{t}-\frac{\tau}{2})g(\bar{t}+\frac{\tau}{2})\,,
\ee  
so
\be
B_i=\int_0^\infty \frac{d\xi}{\pi}\int_{-\infty}^\infty d\tau
F_i(\tau) e^{-i\xi\tau}= \int_0^\infty \frac{d\xi}{\pi}
\hat{F_i}(-\xi)=\frac{1}{\pi}
\int_{-\infty}^\infty d\xi\,\Theta(\xi) \hat{F_i}(-\xi)\,.
\ee
The integral is the distribution $\Theta$ applied to $\hat
F_i(-\xi))$, which is the Fourier transform of $\Theta$ applied to
$F_i(-\tau)$.  The Fourier transform of the $\Theta$ function acts on
a function $f$ as \cite{Gelfand:functions}
\be \label{FTheta}
\Theta[f] = i P \int_{-\infty}^\infty d\tau  
\left(\frac{1}{\tau} f(\tau) \right) + \pi f(0)\,,
\ee
where $P$ denotes principal value, so
\be\label{B46}
B_i=-\frac{i}{\pi} P \int_{-\infty}^\infty d\tau  \left(
\frac{1}{\tau} F_i(\tau) \right) + F_i(0)\,.
\ee

The first of the non-singular terms is a constant: $f_4$ does not
depend on $\tau$.  Thus $F_4$ is even in $\tau$, and only the second term
of Eq.~(\ref{B46}) contributes, giving
\be\label{B4}
B_4=F_4(0)=-\frac{1}{128\pi^2} \int_{-\infty}^{\infty}d\bar{t}\,
g(\bar{t})^2 \left[V_{,tt}(\bar{t}) + V_{,ii}(\bar{t}) \right]\,.
\ee
The functions $f_5$ and $f_6$ are odd in $\tau$, so in these
cases only the first term in Eq.~(\ref{B46}) contributes.
Equations~(\ref{f5},\ref{Fi},\ref{B46}) give
\be\label{B5}
B_5=\frac{1}{256\pi^3}  \int_{-\infty}^\infty d\tau \frac{1}{\tau}
\int_{-\infty}^\infty d\bar{t}\,g(\bar{t}-\frac{\tau}{2})g(\bar{t}+\frac{\tau}{2}) \int d\Omega\,\nabla_{\bar{x}}^2 V^{(1)} (\bar{x}+\frac{|\tau|}{2} \Omega) \sgn{\tau} 
\ee
and Eqs.~(\ref{f6},\ref{Fi},\ref{B46}) give
\be
B_6=-\frac{1}{64\pi^3}  \int_{-\infty}^\infty d\tau \frac{1}{\tau} \int_{-\infty}^\infty d\bar{t}\,g(\bar{t}-\frac{\tau}{2})g(\bar{t}+\frac{\tau}{2}) \int d\Omega\,\partial_{\tau}^2\left[ V^{(3)}(\bar{x}+\frac{|\tau|}{2} \Omega) \sgn{\tau} \right]\,.
\ee
Here we can integrate by parts twice, giving
\be \label{B6}
B_6=-\frac{1}{64\pi^3}  \int_{-\infty}^\infty d\tau
\int_{-\infty}^\infty d\bar t\,
\partial_\tau^2 \left[\frac{1}{\tau} g(\bar{t}-\frac{\tau}{2})g(\bar{t}+\frac{\tau}{2})  \right]
\int d\Omega\,V^{(3)}(\bar{x}+\frac{|\tau|}{2} \Omega) \sgn{\tau}\,.
\ee

\section{The Quantum Inequality}
\label{sec:QI}

Now we have can collect all the terms of $B$ from
Eqs.~(\ref{B7},\ref{B8},\ref{B1},\ref{B2},\ref{B3},\ref{B4},\ref{B5},\ref{B6}).
Since $B_7$ is made of the same quantities as $B_4$, we
merge these together.  We find
\be\label{BforV}
B = \frac{1}{16\pi^2}\left[ I_1
-\frac{1}{2} I_2
-\frac{1}{3} I_3
-\frac{1}{8} I_4
+\frac{1}{16\pi} I_5
-\frac{1}{4\pi} I_6\right]
-I_7\,,
\ee
where
\bml\label{I}\bea
I_1&=&\int_{-\infty}^{\infty} dt\,g''(t)^2 \\
I_2&=&
\int_{-\infty}^\infty d\bar{t}\, V(\bar{t})
[g(\bar{t}) g''(\bar{t}) - g'(\bar{t}) g'(\bar{t})]\\
\label{I3}I_3  &=&\int_{-\infty}^{\infty} d\tau \ln{|\tau|}\sgn{\tau}
\int_{-\infty}^{\infty} d\bar t\,V_{,ii}(\bar{t})
g(\bar{t}-\frac{\tau}{2})g'(\bar{t}+\frac{\tau}{2})\\
\label{I4}I_4 &=& \int_{-\infty}^{\infty}d\bar{t}\,g(\bar{t})^2 
\left[\frac{7}{6}V_{,tt} (\bar{t})+\frac{5}{6}
 V_{,ii}(\bar{t}) \right]
\\
I_5 &=&\int_{-\infty}^{\infty} d\tau \frac{1}{\tau}\int_{-\infty}^\infty d\bar{t}\,
g(\bar{t}-\frac{\tau}{2})g(\bar{t}+\frac{\tau}{2})
\int d\Omega\,\nabla_{\bar{x}}^2 V^{(1)} (\bar{x}+\frac{|\tau|}{2} \Omega)
\sgn\tau
\\
\label{I6}I_6&=&\int_{-\infty}^{\infty} d\tau\int_{-\infty}^\infty d\bar t\,
\partial_\tau^2
\left[\frac{1}{\tau}g(\bar{t}-\frac{\tau}{2})g(\bar{t}+\frac{\tau}{2}) \right]
\int d\Omega\,V^{(3)}(\bar{x}+\frac{|\tau|}{2} \Omega)
\sgn{\tau}\\
\label{I7}I_7 &=&C\int_{-\infty}^{\infty}d\bar{t}\, g(\bar{t})^2 V_{,ii}(\bar{t})\,.
\elea
In Eq.~(\ref{I3}), $\ln|\tau|$ really means $\ln(|\tau|/l)$, where $l$ is
the arbitrary length discussed in Sec.~\ref{sec:FSQEI}.  The choice of
a different length changes Eqs.~(\ref{I3},\ref{I4}) in compensating
ways so that $B$ is unchanged.

Equations~(\ref{QI},\ref{BforV},\ref{I}) give a quantum inequality
useful when the potential $V$ is known and so the integrals in
Eqs.~(\ref{I}) can be done.  If we only know that $V$ and its
derivatives are restricted by the bounds of Eq.~(\ref{Vmax}), then we
can restrict the magnitude of each term of Eq.~(\ref{BforV}) and add
those magnitudes.  We start with
\bea
|I_2|&\le& \int_{-\infty}^\infty d\bar{t} |V(\bar{t})|
[g(\bar{t}) g''(\bar{t}) - g'(\bar{t}) g'(\bar{t})]
\le \Vmax \int_{-\infty}^\infty d\bar{t}\left[g(\bar{t})|g''(\bar{t})|
+  g'(\bar{t})^2\right] \,.
\eea
The cases of $I_3$, $I_4$, and $I_7$ are similar.  For $I_5$ and $I_6$, it is
useful to take explicit forms for the Taylor series remainders.  From
Eq.~(\ref{V1}), we see that
\be
\left|\int d\Omega\,\nabla_{\bar{x}}^2 V^{(1)}
(\bar{x}+\frac{|\tau|}{2} \Omega)\right|
\le \frac{|\tau|}{2}\int d\Omega|\nabla^2 V_{,i}||\Omega^i|
\le \frac{3|\tau|}{2} \Vmax'''\sum_i\int d\Omega|\Omega^i|
= 9\pi|\tau| \Vmax'''\,.
\ee
Similarly from Eq.~(\ref{V3}) we have
\bea
\left|\int d\Omega\,V^{(3)}(\bar{x}+\frac{|\tau|}{2} \Omega)\right|
&\le&\frac{|\tau|^3}{48}\int d\Omega |V_{,ijk}||\Omega^i \Omega^j \Omega^k|\\
&\le& \frac{|\tau|^3}{48}\Vmax'''\sum_{ijk}
\int d\Omega|\Omega^i \Omega^j \Omega^k|\nonumber
=\frac{2\pi+1}{8}|\tau|^3\Vmax'''\,,
\eea
We can then perform the derivatives in Eq.~(\ref{I6}) and take the
absolute value of each resulting term separately.

We define
\bml\label{J17}\bea
J_2&=&\int_{-\infty}^\infty dt\left[g(t)|g''(t)|+g'(t)^2\right]\\
J_3&=&\int_{-\infty}^\infty dt \int_{-\infty}^\infty dt' |g'(t')|g(t) |\!\ln{|t'-t|}| \\ J_4&=&\int_{-\infty}^\infty dt\,g(t)^2\\
J_5&=&\int_{-\infty}^\infty dt \int_{-\infty}^\infty dt' g(t)g(t') \\
J_6&=&\int_{-\infty}^\infty dt \int_{-\infty}^\infty dt' |g'(t')|g(t)|t'-t|\\
J_7&=&\int_{-\infty}^\infty dt \int_{-\infty}^\infty dt' 
\left [g(t)|g''(t')| +g'(t)g'(t')\right] (t'-t)^2
\elea
and find
\blea
|I_2| &\le & \Vmax J_2\\
|I_3| &\le &3\Vmax'' J_3\\
|I_4| &\le & \frac{11}{3}\Vmax''J_4\\
|I_5|&\le &9\pi\Vmax''' J_5\\
|I_6|&\le &\frac{2\pi+1}{16}\Vmax'''\left(4J_5+4J_6+J_7\right)\\
|I_7|&\le & 3|C|\Vmax''J_4\,.
\elea

Thus we have
\bea
\label{final}
\int_{\mathbb{R}} d\tau\,g(t)^2\langle T^{ren}_{00}\rangle_{\omega}
(t,0) \geq- \frac{1}{16\pi^2} \bigg\{& &I_1+\frac12\Vmax J_2
+\Vmax''\left[J_3+\left(\frac{11}{24}+48\pi^2|C|\right) J_4\right]
\nonumber\\
&&+\Vmax''' \left[\frac{11\pi+1}{16\pi}J_5
+\frac{2\pi+1}{64\pi}(4J_6+J_7)\right] \bigg\}\,.
\eea

\subsection{An example for a specific sampling function}

An example of the quantum inequality with a specific sampling function
$g$ is the following. Consider a Gaussian sampling function
\be
g(t)=e^{-t^2/t_0^2}\,,
\ee
where $t_0$ is a positive number with the dimensions of $t$.
Then the integrals of Eqs.~(\ref{J17}), calculated numerically, become
\bea
J_1= 3.75 t_0^{-3}
&& J_2= 3.15 t_0^{-1} \nonumber\\
J_3= 2.70 t_0
&& J_4=1.25 t_0 \\
J_5= 3.14 t_0^2
&& J_6= 3.57 t_0^2\nonumber\\
J_7= 3.58 t_0^2\,,\nonumber
\eea
so the right hand side of Eq.~(\ref{final}) becomes
\be
-\frac{1}{16 \pi^2 t_0^3} \left\{ 3.75+3.15\Vmax t_0^2+(3.26+591.25 |C|) \Vmax'' t_0^4+ 2.86 \Vmax''' t_0^5 \right\}\,.
\ee

\section{Conclusion}
\label{sec:conclusions}

In this work we have demonstrated a quantum inequality for a flat
spacetime with a background potential, considered as a first-order
correction, using a general inequality presented by Fewster and Smith
\cite{Fewster:2007rh}. We calculated the necessary terms from the
Hadamard series and the antisymmetric part of the two-point function
to get $\tilde{H}$. Next we Fourier transformed the terms, which are,
as expected, free of divergences, to derive a bound for a given
background potential.  We then calculated the maximum values of these
terms to give a bound that applies to any potential whose value and
first three derivatives are bounded.

To show the meaning of this result, in the last section we presented
an example for a specific sampling function. By studying the result we
can see the meaning of the right hand side of our quantum
inequality. The first term of the bound goes as $t_0^{-3}$, where
$t_0$ is the sampling time, and agrees with the quantum inequality
with no potential \cite{Eveson:2007ns}. The rest of the terms show the
effects of the potential to first order.  These corrections will
be small, provided that
\blea
\Vmax t_0^2 &\ll& 1\label{Vsmall}\\
\Vmax'' t_0^4&\ll& 1\label{V2small}\\\
\Vmax''' t_0^5&\ll& 1\label{V3small}\,.
\elea

Equation~(\ref{Vsmall}) says that the potential is small when its
effect over the distance $t_0$ is considered.  Given
Eq.~(\ref{Vsmall}), Eqs.~(\ref{V2small},\ref{V3small}) say,
essentially, that the distance over which $V$ varies is large compared
to $t_0$, so that each additional derivative introduces a factor less
than $t_0^{-1}$.

Finally, it is interesting to note the relation of the current work to
the case of a spacetime with bounded curvature. Since the Hadamard
coefficients in that case are components of the Riemann tensor and its
derivatives, we expect that the bound will be the flat space term plus
correction terms that depend on the maximum values of the curvature
and its derivatives, just as in our case they depend on the the
potential and its derivatives. We intend to analyze that case in
future work.
 
\section*{Acknowledgments}

We thank Larry Ford for helpful conversations.  This research was
supported in part by grant RFP3-1014 from The Foundational Questions
Institute (fqxi.org).  E-A.\ K.\ gratefully acknowledges support from
a John F. Burlingame Graduate Fellowship in Physics.

\appendix 

\section{Fourier transforms of some distributions involving logarithms}

In this appendix will compute the Fourier transforms of the
distributions given by
\bea
u(\tau) &=& \ln|\tau|\\
v(\tau) &=& \ln(-\tau_-^2)\,.
\eea
We write $u$ as a distributional limit,
\be
u =  \lim_{\epsilon\to0^+} u_\epsilon\,,
\ee
where
\be
u_\epsilon(\tau) = \ln|\tau|e^{-\epsilon|\tau|}\,,
\ee
so its Fourier transform is
\be
\hat u_\epsilon(k) =\int_{-\infty}^\infty d\tau \ln|\tau|e^{-\epsilon|\tau|}e^{i k \tau}
= 2 \Re\int_0^\infty d\tau \ln \tau \,e^{(ik-\epsilon)\tau}
= -2 \Re\frac{\gamma + \ln(\epsilon-ik)}{\epsilon-ik}\,.
\ee
Thus the action of $\hat u$ on a test function $f$ is
\be
\hat u[f] = -2\lim_{\epsilon\to0^+} \Re\int_{-\infty}^\infty dk
\frac{\gamma + \ln(\epsilon-ik)}{\epsilon-ik} f(k)\,.
\ee
The term involving $\gamma$ is
\be
- 2\gamma \lim_{\epsilon\to0^+} \int_{-\infty}^\infty dk
\frac{\epsilon}{k^2+\epsilon^2}f(k)
= - 2\pi\gamma f(0)\,.
\ee
In the other term we integrate by parts,
\bea
- 2 \lim_{\epsilon\to0^+}\Re \int_{-\infty}^\infty dk
\frac{\ln(\epsilon-ik)}{\epsilon-ik} f(k)
&=& -\lim_{\epsilon\to0^+}\Im\int_{-\infty}^\infty dk\,f'(k) [\ln(\epsilon-ik)]^2\\
&=& -\Im\int_{-\infty}^\infty dk\,f'(k) [\ln|k|-i(\pi/2)\sgn k]^2\\
&=& \pi \int_{-\infty}^\infty dk\,f'(k) \ln|k|\sgn k\,,
\eea
and thus
\be\label{uhat}
\hat u[f] = \pi\int_{-\infty}^\infty dk\,f'(k) \ln|k|\sgn k -2\pi \gamma f(0)\,.
\ee
Since the Fourier transform of the constant $\gamma$ is just
$2\pi\gamma\delta(k)$, the transform of
\be\label{distw}
w(\tau) = \ln|\tau| + \gamma
\ee
is just
\be\label{distwhat}
\hat w[f] = \pi\int_{-\infty}^\infty dk\,f'(k) \ln|k|\sgn k\,.
\ee

Now
\be\label{vdist1}
v(\tau) = \lim_{\epsilon\to0}\ln(-(\tau-i\epsilon)^2) = 2\ln|\tau|+\pi
i \sgn \tau\,.
\ee
The Fourier transform of $\sgn$ acts on $f$ as \cite{Gelfand:functions}
\be\label{sgnhat}
2iP \int_{-\infty}^\infty dk\,\frac{f(k)}{k}
= -2i\int_{-\infty}^\infty dk\,f'(k) \ln |k|\,,
\ee
Putting Eqs.~(\ref{uhat},\ref{sgnhat}) in Eq.~(\ref{vdist1}) gives
\be
\hat v[f] = 4\pi\int_0^\infty dk\,f'(k) \ln|k| -4\pi \gamma f(0)\,.
\ee

\bibliography{no-slac,paper}

\begin{thebibliography}{12}
\expandafter\ifx\csname natexlab\endcsname\relax\def\natexlab#1{#1}\fi
\expandafter\ifx\csname bibnamefont\endcsname\relax
  \def\bibnamefont#1{#1}\fi
\expandafter\ifx\csname bibfnamefont\endcsname\relax
  \def\bibfnamefont#1{#1}\fi
\expandafter\ifx\csname citenamefont\endcsname\relax
  \def\citenamefont#1{#1}\fi
\expandafter\ifx\csname url\endcsname\relax
  \def\url#1{\texttt{#1}}\fi
\expandafter\ifx\csname urlprefix\endcsname\relax\def\urlprefix{URL }\fi
\providecommand{\bibinfo}[2]{#2}
\providecommand{\eprint}[2][]{\url{#2}}

\bibitem[{\citenamefont{Ford}(1978)}]{Ford:1978qya}
\bibinfo{author}{\bibfnamefont{L.}~\bibnamefont{Ford}},
  \bibinfo{journal}{Proceedings of the Royal Society A: Mathematical, Physical
  and Engineering Sciences} \textbf{\bibinfo{volume}{364}},
  \bibinfo{pages}{227} (\bibinfo{year}{1978}).

\bibitem[{\citenamefont{Ford and Roman}(1995)}]{Ford:1994bj}
\bibinfo{author}{\bibfnamefont{L.}~\bibnamefont{Ford}} \bibnamefont{and}
  \bibinfo{author}{\bibfnamefont{T.~A.} \bibnamefont{Roman}},
  \bibinfo{journal}{Phys.Rev.} \textbf{\bibinfo{volume}{D51}},
  \bibinfo{pages}{4277} (\bibinfo{year}{1995}), \eprint{gr-qc/9410043}.

\bibitem[{\citenamefont{Fewster et~al.}(2007)\citenamefont{Fewster, Olum, and
  Pfenning}}]{Fewster:2006uf}
\bibinfo{author}{\bibfnamefont{C.~J.} \bibnamefont{Fewster}},
  \bibinfo{author}{\bibfnamefont{K.~D.} \bibnamefont{Olum}}, \bibnamefont{and}
  \bibinfo{author}{\bibfnamefont{M.~J.} \bibnamefont{Pfenning}},
  \bibinfo{journal}{Phys. Rev.} \textbf{\bibinfo{volume}{D75}},
  \bibinfo{pages}{025007} (\bibinfo{year}{2007}), \eprint{gr-qc/0609007}.

\bibitem[{\citenamefont{Graham and Olum}(2007)}]{Graham:2007va}
\bibinfo{author}{\bibfnamefont{N.}~\bibnamefont{Graham}} \bibnamefont{and}
  \bibinfo{author}{\bibfnamefont{K.~D.} \bibnamefont{Olum}},
  \bibinfo{journal}{Phys. Rev.} \textbf{\bibinfo{volume}{D76}},
  \bibinfo{pages}{064001} (\bibinfo{year}{2007}), \eprint{0705.3193}.

\bibitem[{\citenamefont{Fewster and Roman}(2003)}]{Fewster:2002ne}
\bibinfo{author}{\bibfnamefont{C.~J.} \bibnamefont{Fewster}} \bibnamefont{and}
  \bibinfo{author}{\bibfnamefont{T.~A.} \bibnamefont{Roman}},
  \bibinfo{journal}{Phys.Rev.} \textbf{\bibinfo{volume}{D67}},
  \bibinfo{pages}{044003} (\bibinfo{year}{2003}), \eprint{gr-qc/0209036}.

\bibitem[{\citenamefont{Ford and Roman}(1996)}]{Ford:1995wg}
\bibinfo{author}{\bibfnamefont{L.~H.} \bibnamefont{Ford}} \bibnamefont{and}
  \bibinfo{author}{\bibfnamefont{T.~A.} \bibnamefont{Roman}},
  \bibinfo{journal}{Phys. Rev.} \textbf{\bibinfo{volume}{D53}},
  \bibinfo{pages}{5496} (\bibinfo{year}{1996}), \eprint{gr-qc/9510071}.

\bibitem[{\citenamefont{Kontou and Olum}(2013)}]{Kontou:2012ve}
\bibinfo{author}{\bibfnamefont{E.-A.} \bibnamefont{Kontou}} \bibnamefont{and}
  \bibinfo{author}{\bibfnamefont{K.~D.} \bibnamefont{Olum}},
  \bibinfo{journal}{Phys.Rev.} \textbf{\bibinfo{volume}{D87}},
  \bibinfo{pages}{064009} (\bibinfo{year}{2013}), \eprint{1212.2290}.

\bibitem[{\citenamefont{Fewster and Smith}(2008)}]{Fewster:2007rh}
\bibinfo{author}{\bibfnamefont{C.~J.} \bibnamefont{Fewster}} \bibnamefont{and}
  \bibinfo{author}{\bibfnamefont{C.~J.} \bibnamefont{Smith}},
  \bibinfo{journal}{Annales Henri Poincare} \textbf{\bibinfo{volume}{9}},
  \bibinfo{pages}{425} (\bibinfo{year}{2008}), \eprint{gr-qc/0702056}.

\bibitem[{\citenamefont{Wald}(1994)}]{Wald:qft}
\bibinfo{author}{\bibfnamefont{R.~M.} \bibnamefont{Wald}},
  \emph{\bibinfo{title}{Quantum Field Theory in Curved Spacetime and Black Hole
  Thermodynamics}} (\bibinfo{publisher}{Chicago University Press},
  \bibinfo{year}{1994}).

\bibitem[{\citenamefont{Wald}(1978)}]{Wald:1978pj}
\bibinfo{author}{\bibfnamefont{R.~M.} \bibnamefont{Wald}},
  \bibinfo{journal}{Phys.Rev.} \textbf{\bibinfo{volume}{D17}},
  \bibinfo{pages}{1477} (\bibinfo{year}{1978}).

\bibitem[{\citenamefont{Gel'fand and Shilov}(1964)}]{Gelfand:functions}
\bibinfo{author}{\bibfnamefont{I.~M.} \bibnamefont{Gel'fand}} \bibnamefont{and}
  \bibinfo{author}{\bibfnamefont{G.~E.} \bibnamefont{Shilov}},
  \emph{\bibinfo{title}{Generalized functions}} (\bibinfo{publisher}{Academic
  press}, \bibinfo{address}{New York and London}, \bibinfo{year}{1964}).

\bibitem[{\citenamefont{Eveson and Fewster}(2007)}]{Eveson:2007ns}
\bibinfo{author}{\bibfnamefont{S.~P.} \bibnamefont{Eveson}} \bibnamefont{and}
  \bibinfo{author}{\bibfnamefont{C.~J.} \bibnamefont{Fewster}},
  \bibinfo{journal}{J.Math.Phys.} \textbf{\bibinfo{volume}{48}},
  \bibinfo{pages}{093506} (\bibinfo{year}{2007}), \eprint{math-ph/0702074}.

\end{thebibliography}

\end{document}